\documentclass[preprint,prb,aps,showpacs,showkeys,superscriptaddress]{revtex4-1}

\renewcommand{\Re}[1]{\mathrm{Re}\left(#1\right)}
\renewcommand{\Im}[1]{\mathrm{Im}\left(#1\right)}
\newcommand{\cd}{\hat{c}^{\dagger}}
\newcommand{\cv}{\hat{c}}
\newcommand{\pd}{\mathbf{\hat{p}^{\dagger}}}
\newcommand{\pv}{\mathbf{\hat{p}}}
\newcommand{\pds}[1]{\hat{p}^{\dagger}_{#1}}
\newcommand{\pvs}[1]{\hat{p}^{\phantom{\dagger}}_{#1}}
\newcommand{\bd}{\hat{b}^{\dagger}}
\newcommand{\bv}{\hat{b}}
\newcommand{\avd}{\hat{a}_{\mathrm{v}}^{\dagger}}
\newcommand{\av}{\hat{a}_{\mathrm{v}}^{\phantom{\dagger}}}
\newcommand{\acd}{\hat{a}_{\mathrm{c}}^{\dagger}}
\newcommand{\ac}{\hat{a}_{\mathrm{c}}^{\phantom{\dagger}}}
\newcommand{\avdj}{\hat{a}_{\mathrm{v},j}^{\dagger}}
\newcommand{\avj}{\hat{a}_{\mathrm{v},j}^{\phantom{\dagger}}}
\newcommand{\acdj}{\hat{a}_{\mathrm{c},j}^{\dagger}}
\newcommand{\acj}{\hat{a}_{\mathrm{c},j}^{\phantom{\dagger}}}

\newcommand{\NQD}{N_{\mathrm{QD}}}
\newcommand{\omegacav}{\ensuremath{{\omega_{\mathrm{cav}}}}}
\newcommand{\omegabar}{\ensuremath{{\bar{\omega}}}}
\newcommand{\omegav}{\ensuremath{{\omega_{\mathrm{v}}}}}
\newcommand{\omegac}{\ensuremath{{\omega_{\mathrm{c}}}}}

\newcommand{\omegaL}{\ensuremath{{\omega_{\mathrm{L}}}}}

\newcommand{\Deltasch}{\ensuremath{{\tilde{\Delta}}}}
\newcommand{\gammasch}{\ensuremath{{\tilde{\gamma}}}}

\newcommand{\gammapd}{\ensuremath{{\gamma_{\mathrm{PD}}}}}

\newcommand{\nHF}{n_{\mathrm{HF}}}
\newcommand{\nLF}{n_{\mathrm{LF}}}
\usepackage[utf8]{inputenc}
\usepackage[english]{babel}
\usepackage{amsmath,amssymb,amsfonts,empheq}
\usepackage{microtype}
\usepackage{graphicx,epstopdf}
\usepackage{color}
\usepackage{mathtools}
\usepackage{bbm}

\usepackage{rotating}

\begin{document}

\author{Nicolas L. Naumann}
\author{Leon Droenner}
%\email[Corresponding author:]{naumann@itp.tu-belin.de}
\author{Alexander Carmele}
\author{Andreas Knorr}
\affiliation{Institut f\"ur Theoretische Physik, Nichtlineare Optik und Quantenelektronik, Technische Universit\"at Berlin, Hardenbergstr. 36, 10623 Berlin, Germany}
\author{Weng W. Chow}
\affiliation{Sandia National Laboratories, Albuquerque, New Mexico 87185-1086, USA}
\author{Julia Kabuss}
\affiliation{Institut f\"ur Theoretische Physik, Nichtlineare Optik und Quantenelektronik, Technische Universit\"at Berlin, Hardenbergstr. 36, 10623 Berlin, Germany}

\title{Solid state based analog of optomechanics}

\begin{abstract}
We investigate a semiconductor quantum dot as a microscopic analog
of a basic optomechanical setup.
We show, that optomechanical features can be reproduced by the solid-state 
platform, arising from parallels of the
underlying interaction processes, which in the optomechanical case is the 
radiation pressure coupling and in the
semiconductor case the electron-phonon coupling. In optomechanics, phonons 
are typically induced via confined
photons, acting on a movable mirror, while in the semiconductor system the 
phonons are emitted by the laser-driven electronic system.
There are analogous effects present for both 
systems, featuring bistabilities, optically induced phonon lasing and enhanced phonon loss. 
Nonetheless, the different statistical nature of the optical cavity and the electronic system
also leads to qualitative differences.
\end{abstract}

\maketitle
\date{\today}
\selectlanguage{english}

\section{Introduction}
Quantum optomechanics (OM) has been rapidly developing into a major research area 
over the last decades \cite{omrevasp,KippenbergScience08}.
Intense activities are fueled by interesting physics based on radiation 
pressure 
arising from momentum transfer between photons and optical
surfaces. The phenomenon affects ultra-accurate measurements (e.g., 
gravitational wave detection \cite{PhysRevLett.45.75})
and can lead to nanoscale motors and refrigerators 
\cite{PhysRevLett.112.150602}.

This manuscript proposes and analyzes a different physical system for exploring a 
semiconductor (SC) analog of quantum optomechanics.
It differs from the atomic-molecular-optical (AMO) approach to optomechanics in 
that radiation pressure is replaced by the electron-phonon interaction. The 
presence of electron-phonon interaction usually leads to heating in 
optoelectronic devices \cite{0953-8984-13-32-312}.
However, if optically addressing phonon assisted resonances or when properly 
modifying and harnessing electron-photon interaction in
an acoustic cavity, the same combination of interactions can be useful, e.g., 
provide high-temperature single-photon generation \cite{SinglePhotonReview},
optoelectronic self-cooling and memory in silicon photonics.% \cite{SiliconPhotonics4063407}. There is 
There is interesting physics as well, e.g. enabling of
strong-coupling physics, via a quantum-mechanical polariton-polaron 
superposition \cite{restrepo:prl:2014,PhysRevB.87.041305}.\\
Several proposals of phonon cavities
coupling an acoustic phonon mode to the electronic degrees of freedom have been 
made \cite{PhysRevB.75.024301,Soykal11}
and even though there are challenges such as anharmonic phonon decay or 
fluctuations in layer thickness, high-Q phonon
cavities have been realized experimentally 
\cite{PhysRevLett.89.227402,PhysRevLett.102.015502,PhysRevLett.110.037403,LanzillottiKimura201580}.
Besides such a designed electron-acoustic phonon coupling, that enables the 
interaction with a selected acoustic phonon mode, many semiconductor 
nanostructures naturally inherit a strong effective single mode electron-phonon
coupling mechanism, i.e. the coupling to longitudinal optical (LO) phonons \cite{PhysRevLett.98.187401,kabuss:prb:2011}. At 
optical excitation, these systems may find applications as phonon lasers 
\cite{JuliaPhononLaser,*PhysRevB.88.064305}.\\ %,Khurgin ?
A possibility of these new applications
and physical effects comes from the semiconductor electron-phonon coupling 
mechanism being stronger than radiation pressure. Further, the solid 
state platform typically involves much larger phonon frequencies \cite{omrevasp,JuliaPhononLaser,*PhysRevB.88.064305},
especially in the case of LO phonons. This in particular enables a controlled 
selection of subsystems for effective phonon lasing or damping only by the 
choice of optical input frequencies. %Zitat?
\begin{figure}[h!]
\includegraphics[width=0.7\linewidth]{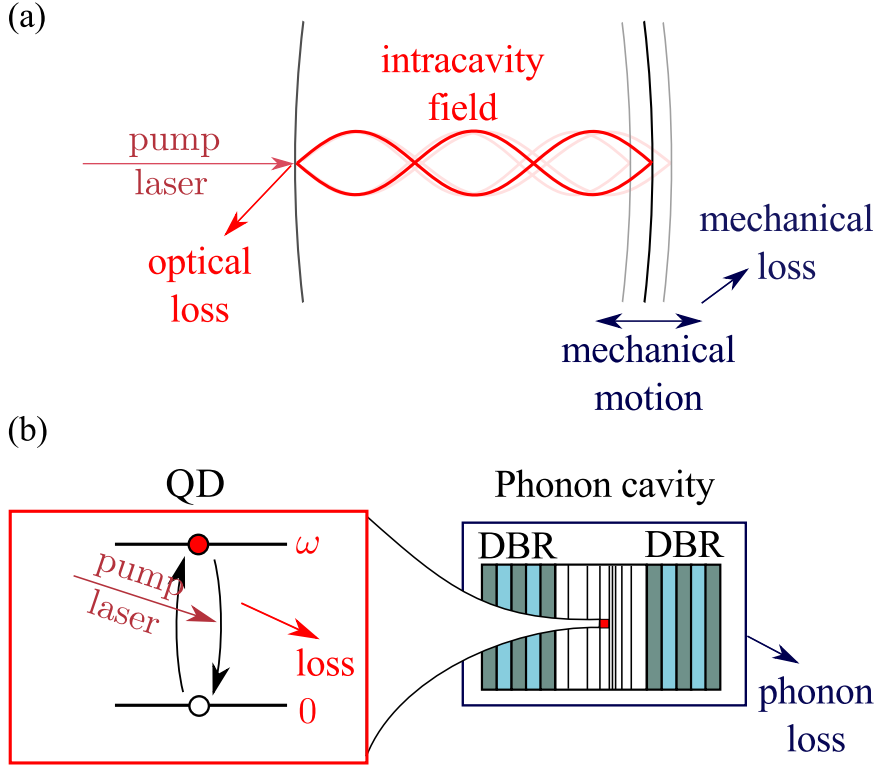}
\caption{\label{fig:setup} (a) Basic setup for investigating cavity 
quantum optomechanics.
(b) Proposed semiconductor quantum optical approach to exploring acoustic 
phonon 
dynamics. The phonon cavity consists of materials of alternating acoustic 
impedance in analogy to an optical distributed Bragg reflectors 
(DBR)\cite{PhysRevLett.89.227402,LanzillottiKimura201580}.}
\end{figure}
The sketch in Fig. \ref{fig:setup}(a) depicts the basic cavity quantum 
optomechanics setup. Photons give momentum to the cavity mirrors upon 
reflection, which then changes  the mirror position and affects ultra-high 
precision (atomic scale) measurements. The Hamiltonian has the form \cite{omrevasp}
\begin{equation}
H_{\mathrm{OM}}=\hbar\omegacav \cd\cv+\hbar\Omega_{\mathrm{m}} \bd\bv+i\hbar E_1 \left(\cd e^{-i\omegaL t} - \cv e^{i\omegaL t} \right)-\frac{\hbar\omega_{\mathrm{cav}}}{L} 
\sqrt{\frac{\hbar}{m\Omega_{\mathrm{m}}}}\cd\cv \frac{\left(\bv+\bd \right)}{\sqrt{2}}. %\underbrace{\cd\cv}_{\text{Photons}} 
%\overbrace{\frac{\left(\bv+\bd \right)}{\sqrt{2}}}^{\mathclap{\text{Mirror displacement}}}
\label{eq:H_OM} 
\end{equation}
Here, $\cv$ ($\cd$) is the cavity phonon annihilation (creation) operator and $\omegacav$ the cavity frequency. $\Omega_{\mathrm{m}}$ is the frequency of the mechanical oscillator and $\bv$ ($\bd$) the corresponding annihilation (creation) operator. The third term models the driving by an external laser and the last term is the coupling due to radiation pressure, where $L$ is the length of the cavity, $m$ the mass of the mirror.\\
The sketch in Fig. \ref{fig:setup}(b) is our proposed alternate experimental 
configuration for the semiconductor system, with the Hamiltonian
\begin{multline}
H_{\mathrm{SC}}=\hbar\omegav\avd\av+\hbar\omegac\acd\ac\\+i\hbar E_1 \left(\acd\av e^{-i\omegaL t} - \avd\ac e^{i\omegaL t} \right)\\+\hbar\sum_{\mathbf{q}} \omega_{\mathbf{q}} \hat{b}_{\mathbf{q}}^{\dagger} \hat{b}_{\mathbf{q}}+\left(g_{\mathbf{q}}^{\mathrm{c}}- g_{\mathbf{q}}^{\mathrm{v}} \right)\acd\ac \hat{b}_{\mathbf{q}}+H.c.
\label{eq:H_SC} 
\end{multline}
Here, $\av$ or $\ac$ ($\avd$ or $\acd$) are the annihilation (creation) operators of an electron in the valance band or conduction band, respectively. In this case, $\hat{b}_{\mathbf{q}}$ ($\hat{b}_{\mathbf{q}}^{\dagger}$) is the annihilation (creation) of a phonon with the frequency $\omega_{\mathbf{q}}$. In this case, the electronic transition is driven optically. Furthermore, the phonon modes will be treated as one effective phonon mode.\cite{PhysRevLett.98.187401,kabuss:prb:2011}\\
We want to emphasize some formal similarities of the Hamiltonians: In place of 
the photon number operator in the AMO case, there is the electron population 
operator. In conventional quantum optomechanics, the photons create or remove 
mirror motion. In our case [Eq.\eqref{eq:H_SC}], a laser-driven electronic transition creates or
annihilates a phonon. The coupling constants are of course different.\\
The Hamiltonians for both systems can be brought to the form of Eq. \eqref{eq:hamiltonian}, where we included additional electronic transitions, which are excited optically, for additional quantum dots.
We now introduce a common nomenclature for the elements of both systems for 
notational convenience and to emphasize the similar roles the components play for the 
effects studied below. Thus, in the optomechanical system the photon mode is of 
high frequency (HF), while in the semiconductor system a two level system,
which models a quantum dot, constitutes the HF part. The low frequency (LF) 
component is the mechanical mode of the oscillating mirror in the 
optomechanical case, whereas in the semicondutor case the phonon
mode inside an acoustic cavity, as discussed before, is the LF 
part.\\
From the sketches [cf. Fig. \ref{fig:setup}], it may seem like the 
semiconductor  version is more complicated than the AMO one. However, with the 
actual experimental setups, the exact opposite is true. Conventional quantum 
optomechanics can involve large and complicated setups, like the LIGO for 
detecting gravitational wave. But there are also microscopic optomechanical
devices, which maximize the optomechanical coupling strength by small oscillator
masses. To do this, a broad range of designs exists in addition to the classical
Fabry-Perot-Setup, using e.g. wispering gallery modes or photonic crystals \cite{omrevasp}.\\
The semiconductor system is on a chip, and very 
much like a photonic integrated circuit with the possibility of incoherent 
pumping of the semiconductor active medium. From the AMO side, connecting the 
two approaches increases enormously the relevance of quantum optomechanics, 
because of the widespread use of optoelectronic devices. For example basic 
quantum mechanical features, such as the formation of mechanical polarons in 
polariton optomechanics \cite{restrepo:prl:2014} as optomechnical counterpart
to the polaron formation in typical semiconductor cavity quantum 
electrodynamics can be expected in quantum hybrid systems. From the 
semiconductor side, making the connection to AMO quantum optomechanics gives 
guidance on what may be interesting or useful to explore, transferring 
predictions and techniques for optomechanics devices onto optoelectronic 
platforms.\\
%However, in the semiconductor case, the optomechanics setup is 
%automatically incorporated via the electron-phonon interaction.
With this guidance, we developed an analogy between the optomechanical coupling and 
semiconductor quantum-dot system and choose as demonstration of similarity 
between the two  approaches three exemplary effects, which can be found in both cases.
The manuscript is structured as follows.
In Sec. \ref{sec:comp}, we introduce the Hamiltonian, which can be brought to
the same form for both systems.
The first effect we consider in Sec. \ref{sec:bistab} is the bistability as a prototypical nonlinear effect.
Next, mechanical lasing, also known as phonon lasing, is discussed in Sec. \ref{sec:lasing}.
It is present in optomechanical \cite{PhysRevLett.96.103901,1367-2630-10-9-095013,omrevasp} as 
well as in the semiconductor systems \cite{JuliaPhononLaser,*PhysRevB.88.064305}. %Khurgin?
Finally, in Sec. \ref{sec:damping}, the basis of laser cooling of mechanical motion will be discussed,
which can be observed in optomechanical systems \cite{naturelasercooling}. Theoretically, it can be 
understood as an enhancement of the effective damping of the mechanical 
oscillator \cite{PhysRevA.77.033804}. We compare the realization of these effects in the above systems.
%the highly nonlinear phenomenon of optical 
%bistability. Furthermore, we show that in both systems the interaction between 
%the high frequency (HF) component and the low frequency 
%(LF) component allows the damping to be strongly enhanced or diminished, 
%depending on the detuning between the resonance of the HF part and the pump 
%laser. In the manuscript we will always show the steps for each of the systems, that 
%is the optomechanical (OM) and the semiconductor (SC) one. This will be done
%in the following manner: At first, in Sec. \ref{sec:comp}, we will introduce the 
%model Hamiltonians of both systems and the resulting semiclassical equations of 
%motion. In section \ref{sec:bistab}, the stationary, nonlinear effect of 
%bistability is presented as an effect present in both systems. Following this 
%we consider two related dynamical effects on the LF component. In Sec. 
%\ref{sec:damping}, we consider the enhancement of damping and in Sec. 
%\ref{sec:lasing} phonon lasing is examined.
%\newpage

\section{Comparison}
\label{sec:comp}

We compare two different physical systems to explore 
mutual effects but also to elaborate on differences. On the one hand, we 
are interested in an optomechanical system, which consists of an optically 
pumped cavity mode, which is coupled by radiation pressure to a mechanically 
oscillating mirror. As a semiconductor counterpart of such an optomechanical 
setup we investigate optically pumped quantum 
dots (QDs), that are coupled to a single phonon mode.\\
The second case may be realized in two ways: The phonon mode is either a 
single confined acoustic phonon mode or an effective mode describing the 
collective motion of several optical phonon modes.\\
Both systems can be described 
by very similar sets of equations of motion, which will be shown in the following.
Thus, even though the main difference is the statistics of the HF mode,
some differences will only arise from the choice of parameters.\\
Despite all of the above mentioned effects being present in both, the OM 
and SC system, their origin is of different nature and thus also their experimental
realization. As we will see in the following, the crucial difference between 
the systems is the behavior of the component, that is coupled to the 
mechanical degree of freedom. In the OM, the mechanical mode 
is coupled to an optical mode, which is bosonic and can thus have an arbitrary 
number of excitations. In the SC system, the phonon mode is coupled 
to one or several QDs, which constitute an electronic and thus a 
fermionic system.\\
In this manuscript the QDs are modeled as a number of two-level systems 
(TLS). Therefore, the SC system can only contain a limited number of 
excitations. But also in more complex electronic systems the number of 
excitations is in principle restricted, which is not the case for the optical cavity
in the OM system. This will lead to some major differences 
between the systems. E.g. cycling processes, discussed in the context of lasing, can be
affected more strongly by electronic lifetimes. However, we will show that within certain parameter 
regimes and a selection of quantities the SC system can approach the OM one 
for a large number of QDs. Then the possible number of excitations becomes very 
large, as in the OM case, which would strictly be the limit of an infinite 
number of QDs.\\

\textit{Hamiltonian ---} The basis of our analysis constitutes a Hamiltonian 
model. The Hamiltonians can be written in the same form for both systems 
\cite{mmlaw,JuliaPhononLaser,*PhysRevB.88.064305}
\begin{equation}
H=\hbar\Omega \bd\bv+\hbar\omega \pd\pv + \hbar g \pd\pv \left(\bv+\bd \right) 
 + i \hbar \mathbf{E} \left(\pd  e^{-i \omegaL t} -\pv e^{ i\omegaL t}\right).
\label{eq:hamiltonian}
\end{equation}
The terms can be identified with the different components of each system: The 
first term is the mechanical or the phonon mode in the optomechanical and the 
semiconductor system, respectively. They are modeled as a
single harmonic oscillator with frequency $\Omega$. Then $\bv$ ($\bd$) is the 
annihilation (creation) operator for one phonon (we will also call the 
mechanical excitations of the oscillating mirror phonons), so the commutation 
relation reads $[\bv,\bd]=1$ in both cases. This constitutes the part of the 
system with low frequency (LF) in comparison to optical frequencies.
The second term models the cavity mode or the QDs. As shorthand notation, we 
introduce the vector $\pv$ with elements $\pvs{j}$, whose length depends on the 
system. In the OM system, there is only one element, namely the optical cavity, 
and thus $\hat{p}_1=\cv$. The cavity is modeled as a harmonic oscillator
oscillating at frequency $\omega$. The annihilation (creation) operator for the 
cavity mode is $\cv$ ($\cd$) (it is also $[\cv,\cd]=1$). The commutation 
relations for the degrees of freedom of the QDs differ due to their 
fermionic nature. In this case, $\pv$ is now a vector with $\NQD$ elements 
$\pvs{j}$, where $\NQD$ is the number of QDs coupled to the mechanical mode. 
In the following, we are interested in the case of $\NQD$ identical QDs
(inhomogeneous broadening is neglected).\\
We can now identify $\hat{p}_j=\avdj\acj$ 
($\hat{p}^{\dagger}_j=\acdj\avj$), with the anti-commutation relations $\lbrace \hat{a}_{\mu,j}^{\phantom{\dagger}},\hat{a}_{\nu,l}^{\dagger}\rbrace =\delta_{\mu,\nu}\delta_{j,l}$.
Here, the $\hat{a}_{\mu,j}^{\phantom{\dagger}}$ ($\hat{a}_{\mu,j}^{\dagger}$) are
the annihilation (creation) operator for an electron in the band $\mu$ of the $j$th
QD mentioned in the introduction. 
of the $j$th QD. Here, we see that either the optical cavity or the QDs constitute
the high frequency (HF) part 
of the respective system, since they oscillate at optical frequencies. The 
third term gives the coupling of the two elements with the coupling strength 
$g$. The number of excitations in the system $\pd\pv$ is coupled to the 
position of the LF mode $q=(\bv+\bd)/\sqrt{2}$. The last term is the pumping 
term, which describes the external pumping of the coherence $\pv$ by
a classical field with frequency $\omegaL$. Here, $\mathbf{E}$ is again a 
vector and consists of the individual pump strengths $E_j$ of the QDs.\\
%
%We can now identify $\hat{p}_j=\sigma_{-,j}$ 
%($\hat{p}^{\dagger}_j=\sigma_{+,j}$), with the commutation relations 
%$[\sigma_{-,j},\sigma_{+,l}]=-\sigma_{z,j} \delta_{l,j}$ and
%$[\sigma_{\pm,j},\sigma_{z,l}]=\mp 2\sigma_{\pm,j} \delta_{l,j}$, where 
%$\sigma_{-,j}$, $\sigma_{+,j}$ and $\sigma_{z,j}$ are the Pauli spin matrices 
%of the $j$th QD. Here, we see that either the optical cavity or the QDs constitute
%the high frequency (HF) part 
%of the respective system, since they oscillate at optical frequencies. The 
%third term gives the coupling of the two elements with the coupling strength 
%$g$. The number of excitations in the system $\pd\pv$ is coupled to the 
%position of the LF mode $q=(\bv+\bd)/\sqrt{2}$. The last term is the pumping 
%term, which describes the external pumping of the coherence $\pv$ by
%a classical field with frequency $\omegaL$. Here, $\mathbf{E}$ is again a 
%vector and consists of the individual pump strengths $E_j$ of the QDs.\\

\textit{Equations of motion ---} From the Hamiltonian \eqref{eq:hamiltonian} 
the equations of motion can be derived.
For the expectation value of some operator $\hat{O}$ they are given by
\begin{align}
\frac{d}{dt}\langle \hat{O} \rangle &= \frac{i}{\hbar} \langle [H,\hat{O}] 
\rangle + \gamma \text{Tr}(\hat{O} \mathcal{L}[\pv]\rho)+ \kappa 
\text{Tr}(\hat{O} \mathcal{L}[\bv]\rho)
\label{eq:ehrenfest}\\
& + \frac{\gammapd}{2} \text{Tr} (\hat{O}\mathcal{L}[\sigma_z]\rho),
\nonumber\\
\mathcal{L}[\hat{X}] \rho &= 2 \hat{X} \rho \hat{X}^{\dagger} - 
\hat{X}^{\dagger}\hat{X}\rho- \rho \hat{X}^{\dagger} \hat{X},
\nonumber
\end{align}
where damping and pure dephasing are taken into account in the Lindblad form.
In the OM system the cavity mode is radiatively damped with $\gamma$, which is 
caused by scattering to other modes, outcoupling and imperfections of the 
mirrors. The mirror is damped with $\kappa$ because of relaxation to the 
environment. In the SC system the damping rate of the excited electronic state 
is given by $2 \gamma$ and the damping rate of the phonon mode is $\kappa$. 
These originate from couplings to not explicitly treated photon and phonon bath 
modes. The pure dephasing exists only for the SC system with
optical phonons.\\
From Eq. \eqref{eq:ehrenfest}, a system of equations is derived. For simplicity,
we assume coherent fields and factorize products of operators into products of
expectations values. Finally, the excitation of the QDs is assumed to be equal for
all QDs and that they follow the same dynamics. Then, the dynamics of
each QD is described by the same equations. In the OM case, there are two
equations, while we need three for the description of the SC system. The 
additional equation occurs since, for the optical cavity system, the number 
of excitations can be factorized into the coherences: $\langle 
\cd\cv\rangle=\langle \cd\rangle\langle\cv\rangle$. In contrast, $\langle 
\sigma_{+,j}\sigma_{-,j}\rangle$ cannot be factorized,
as this would imply that one individual QD can be excited more than once.
After transforming the HF component to a frame rotating at the frequency of
the external pump laser $\omegaL$, the equations of motion read (shorthand
notations for the expectation values: $P= \langle \pvs{1} \rangle e^{i 
\omega_{\mathrm{L}}t}$, $B = \langle \bv\rangle $,
and, only for the SC system, $U = \langle \pds{1}\pvs{1} \rangle$)
\begin{subequations}
\label{eq:eom}
\begin{align}
\dot B &=-(i \Omega +\kappa) B - i g N U
\label{eq:eomb} \\
\dot P &= \left(i\Delta -\gamma-\gammapd\right) P - i g (B+B^*) P
\label{eq:eomp}\\
& + E_1 \left( 1 - U \mp U \right)
\nonumber
\\[3mm]
\dot U &= E_1 (P+P^*) - 2 \gamma U.
\label{eq:eomu}
\end{align}
\end{subequations}
Here, the negative sign in Eq. \eqref{eq:eomp} is valid for the SC system, while the positive one holds
for the OM system. Then, pump term does not depend on the occupation on the HF component, and Eq.
\eqref{eq:eomu} becomes obsolete. Also, the occupation $U=P^*P$.
In the SC case $N=\NQD$, while in the OM case $N=1$,
since only one cavity mode is considered.\\
The equations for the coherences $P$ 
and $B$ are formally identical: The HF component rotates at the frequency
given by the detuning from the external pump laser $\Delta=\omegaL-\omega$ and 
is damped by $\gamma$. The coherence of the HF component is coupled to the 
position of the LF component, which effectively leads to a shifted frequency. 
Additionally, it is pumped by the external laser. In the OM system, the pumping 
term is not limited, while the pump in the SC system is restricted, since the 
QDs are saturated at some point. Finally, the coherence is coupled to the 
excitation of the HF component. The last equation describes the excitation
of the upper level to account for the fermionic behavior in the SC system and
contains a term with pumping and a term with damping.\\
As shorthand notation we define the expectation value for the number of 
excitations in the HF component as $\nHF=N U$, which becomes for the OM system 
$\nHF=P^* P$. In the OM system this is the number of cavity photons, while it 
is the number of electronic excitations in the SC system. The expectation value 
for the number of excitations in the LF component, i.e. the phonon number, is 
defined as $\nLF=B^*B$. The OM equations are also valid for the SC system, when 
an individual quantum dot is excited marginally, which was shown in the context 
of spin systems, modeled by TLSs \cite{PhysRev.58.1098}.

\section{Bistability}
\label{sec:bistab}

\begin{figure}
  \centering
  \includegraphics[width=0.7\linewidth]{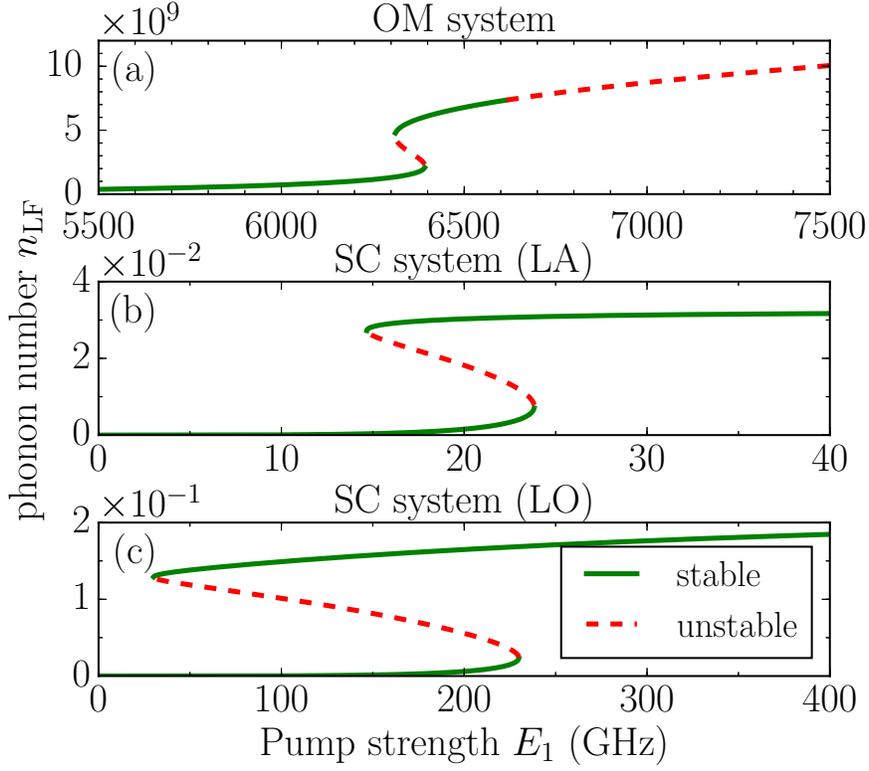}
  \caption{Stationary states of the phonon number $\nLF$ for (a) the OM system, 
(b) the SC system with acoustic phonons, and (c) for the SC system with optical 
phonons. Note the difference in pump strength for bistability to occur.
The parameters are given in Tab. \ref{tab:bistpars}.}
  \label{fig:bistabs}
\end{figure}

\begin{figure}
  \centering
  \includegraphics[width=0.7\linewidth]{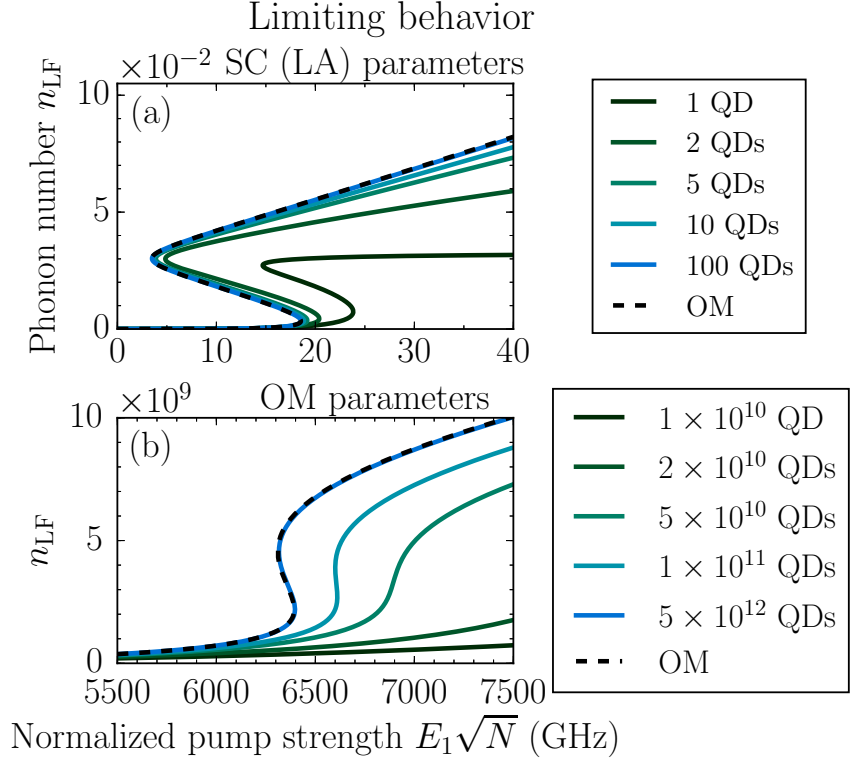}
  \caption{The limiting behavior of the stationary states of the phonon number 
$\nHF$ for the SC system (a) with parameters for LA phonons and (b) with OM 
parameters. An increasing number of TLSs is shown: In the limit 
$N\rightarrow\infty$, the OM solution (dashed) is approached (parameters as in 
\ref{fig:bistabs}(b)).}
  \label{fig:NQDLimit}
\end{figure}

The first example we will give to show the similarities of the systems is the
bistable behavior. The optomechanical bistability has been considered.
\cite{qmombistab} The equation for the steady state can be derived from 
\eqref{eq:eom} by setting the derivatives to zero. Furthermore, the stability of 
the steady state can be investigated by linearizing the system of equations in 
vicinity of the steady state \cite{PaperOptoControl2014}.
The bistability in the phonon number is given in Fig. \ref{fig:bistabs}(a) for 
the OM system and in Figs. \ref{fig:bistabs}(b) and \ref{fig:bistabs}(c) for 
the SC system with acoustic phonons and with optical phonons, respectively.
In all cases a bistability can be observed. For understanding the origin of the 
bistability, we will have a closer look at the forces on the LF component, 
which are given by $F_{\mathrm{tot}}=(\dot{B}-\dot{B}^*)/(\sqrt{2} i)$.\\
There are two competing forces acting on the LF component: The restoring force 
resulting from the harmonic potential $F_h$ on the one hand and the force given 
by the coupling to the HF component $F_c$. In the steady state, an equilibrium 
of forces $F_{\mathrm{tot}}=F_h+F_c=0$ must prevail. When referring to a steady 
state value rather than the variable, we will add the index $s$. In the steady 
state, the restoring force is the same in both cases and linear with respect to 
the displacement $q_s=(B_s+B_s^*)/\sqrt{2}$ of the LF component 
\begin{equation}
F_h=\left(\Omega +\frac{\kappa^2}{\Omega}\right) q_s.
\label{eq:harmonic}
\end{equation}
In contrast to this the force $F_c$, originating from the coupling of the phonon
mode to the electronic excitation, is a nonlinear function of the displacement 
of the LF component
\begin{equation}
F_{\mathrm{c,SC}} = \frac{N g}{\sqrt{2}} 
\frac{1}{1+\frac{\gamma}{\gamma+\gammapd}\frac{\left( \Delta - \sqrt{2} g q_s 
\right)^2+\left(\gamma+\gammapd\right)^2}{2 E_1^2}}.
\label{eq:FcSC}
\end{equation}
The radiation pressure force in the OM case is obtained in the limit $N=1$, $\gammapd=0$ and $E_1^2\ll\left( \Delta-\sqrt{2} g q_s \right)^2+\gamma^2$ 
as
\begin{equation}
F_{\mathrm{c,OM}} = \frac{g}{\sqrt{2}}\frac{2 E_1^2}{\left( \Delta-\sqrt{2} g
q_s \right)^2+\gamma^2}.
\label{eq:FcOM}
\end{equation}
A bistability can only be observed, for detunings $\Delta<0$, because there lies the
maximum of $F_c$, regardless of the sign of $g$. Then, up to three 
equilibrium positions may occur.\\
The difference between the OM and the SC case originates from the fact, that 
the force acting on the phonon mode is affected by the number of excitations in 
the HF component. The radiation pressure force can be increased by increasing $E_1$,
even if the OM coupling constant is small.
For the SC system, the absolute value of $F_{\mathrm{c,SC}}$ is at maximum $\frac{N g}{\sqrt{2}}$. 
Thus, either the coupling constant $g$ has to be large enough, or the number of 
QDs has to be increased. Otherwise, pumping the QDs more strongly does not increase 
the force acting on the phonons, so that it may not become large enough to 
exhibit a bistability. This difference can be traced back to the statistics of the HF component,
since its excitation can be increased arbitrarily by pumping in the OM case,
while it is limited too in the fermionic case.\\
In App. \ref{app:NQDlimit}, we show that the equation for the SC system approaches
the OM one for $N \rightarrow \infty$ and $\gammapd=0$, when the pump strength is rescaled
according to $E_1 \sqrt{\NQD}$. This is illustrated for the SC system with
LA phonons in Fig. \ref{fig:NQDLimit}(a). When the number of QDs is much larger than
the typical number of phonons, the OM solution is reached by the SC system.
In this case the coupling constant
$g$ is large enough, so that already for $N=1$ a bistability can be observed.\\
The same transition is shown for OM parameters in Fig. \ref{fig:NQDLimit}(b).
In this case there is only a bistability for a large number of TLS,
since the coupling constant $g$ is rather small.
For both cases the ratio $\NQD/\nLF \approx 10^3$ gives an estimate for the number of QDs,
for which the SC system coincides with the OM one. Here, $\nLF$ is the stationary phonon number in the regime of the bistability.\\
With this, the major difference in Figs. \ref{fig:bistabs} can be explained: In 
the OM case the phonon number is very high (in the order of $10^9$), while it 
is rather low in the SC systems, since a bistability with a small coupling 
constant can only be observed, if the radiation pressure force is increased 
sufficiently through a high intensity in the optical cavity. In the SC systems, 
the bistability occurs already for very few phonons, since the electron-phonon 
coupling is stronger.\\
Note, that in the SC system with optical phonons also pure dephasing is taken
into account. Then the system is in a regime, where for one QD no bistability can
be observed, but ten QDs suffice to obtain a bistability [cf. Fig. 
\ref{fig:bistabs}(c)].
With the additional pure dephasing, the limiting behavior shown above is not valid.\\
After considering the bistability, which is a shared stationary property of
the systems and motivates the OM system as a limiting case of the SC system
with a large number of QDs, we will consider dynamical effects in the following 
sections.

\section{Lasing}
\label{sec:lasing}

\begin{figure}
  \centering
  \includegraphics[width=0.6\linewidth]{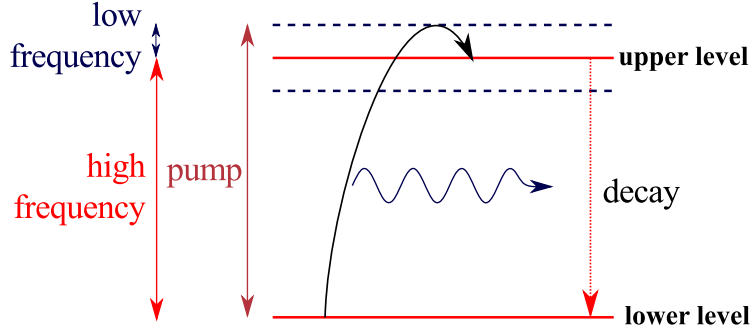}
  \caption{Scheme of the mechanism leading to lasing. See text for explanation.}
  \label{fig:Schemaheat}
\end{figure}

\begin{figure}
  \centering
  \includegraphics[width=0.7\linewidth]{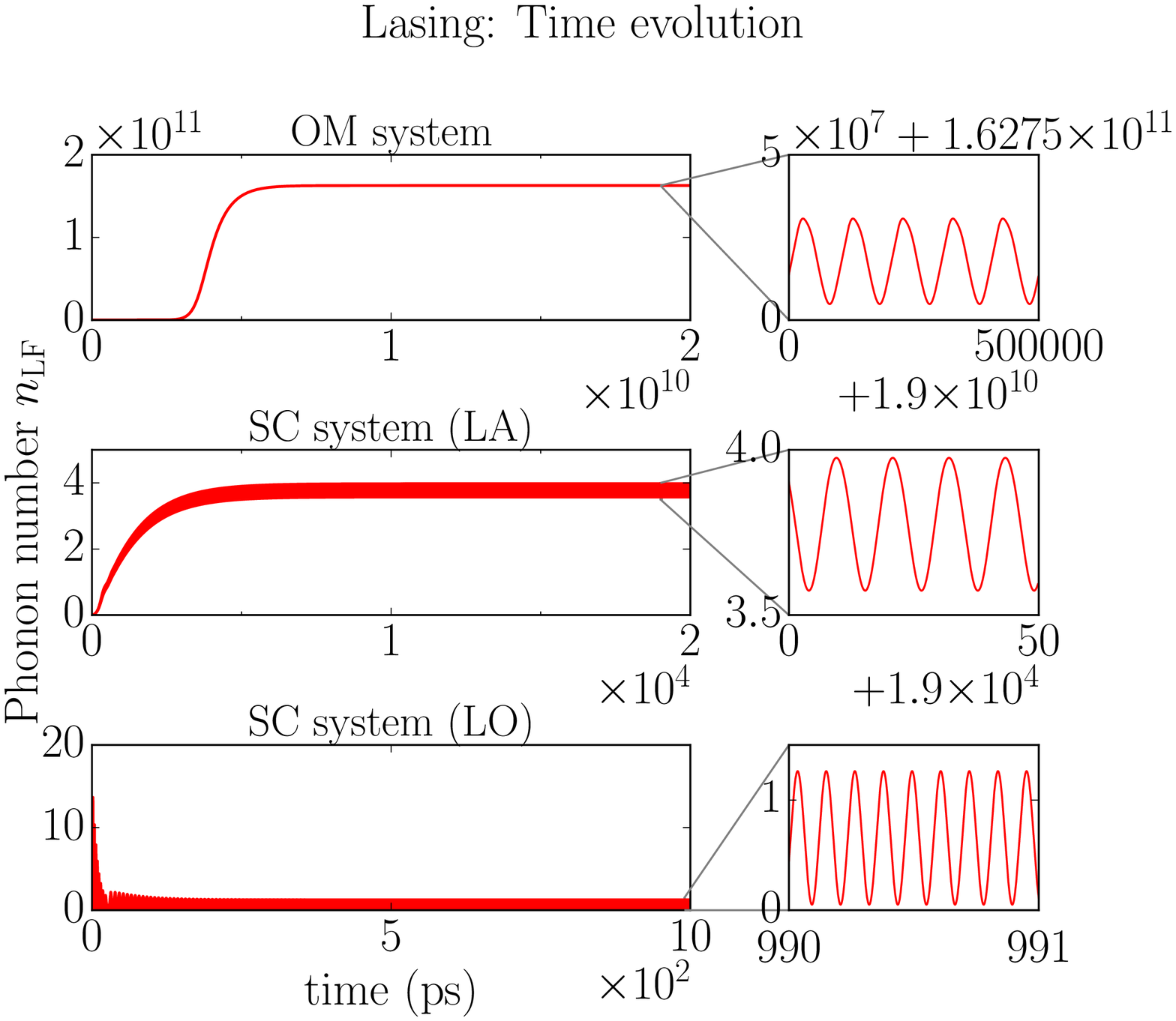}
  \caption{The time evolution for exemplary parameters [cf. Tab. \ref{tab:damppars}] is shown for the OM (a), the SC system with LA phonons (b), and for the SC system with LO phonons.}
  \label{fig:LasingEvoIns}
\end{figure}

\begin{figure}
  \centering
  \includegraphics[width=0.7\linewidth]{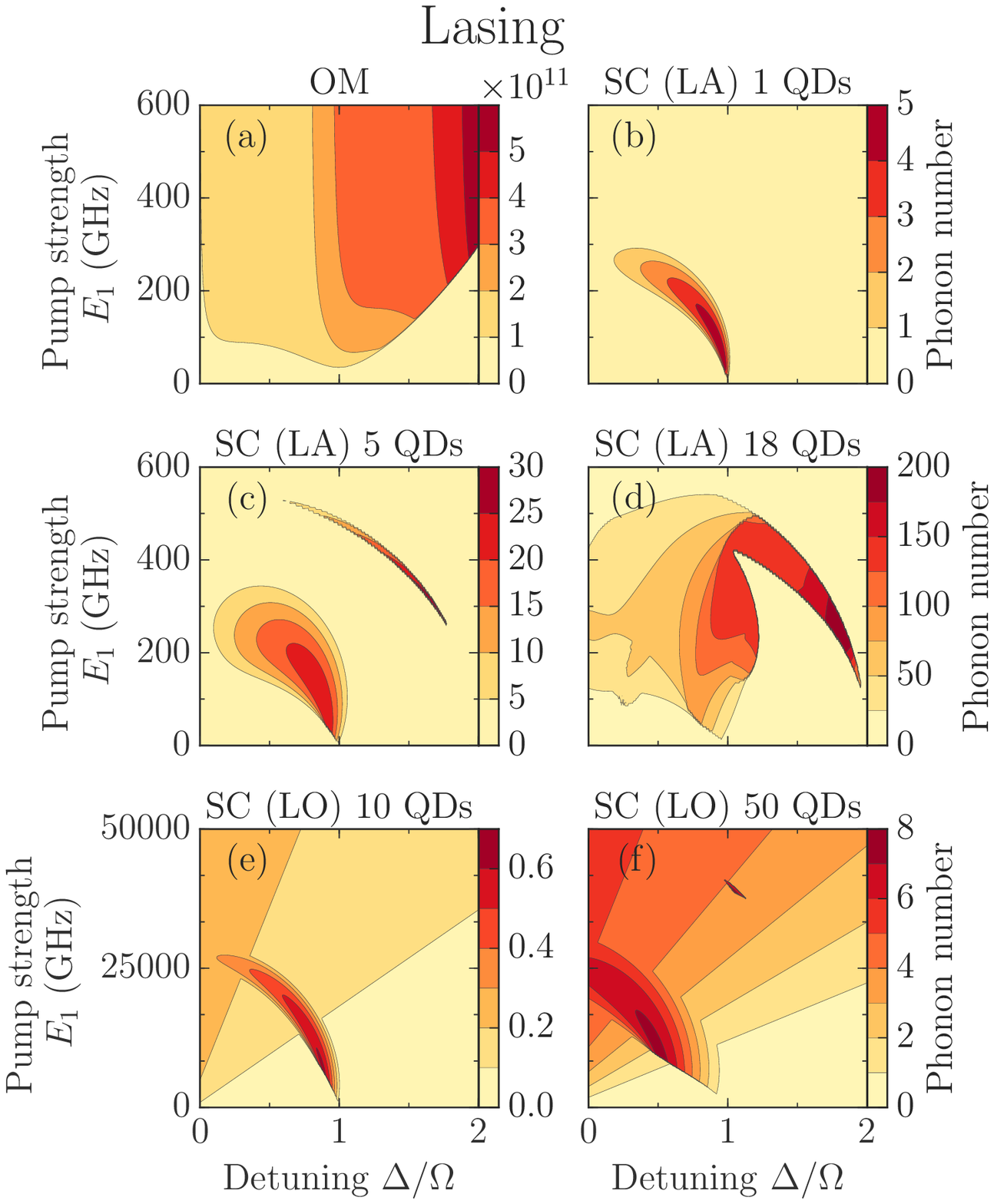}
  \caption{The mean phonon number is plotted as a function of the pump strength and the detuning for the OM system (a), the SC (LA) system with 1 (b), 5 (c), and 18 (d) QDs, and the SC (LO) system with 10 (e) and 50 (f) QDs.}
  \label{fig:Lasingdensity}
\end{figure}

Fig. \ref{fig:NQDLimit} suggests that the differences between the systems are
more quantitative in nature. In the following, we will consider phonon lasing
and damping. These processes, will reveal more profound differences between the two setups.
At first, we will examine the effect known as mechanical or phonon 
lasing: A high number of coherent phonons is created in the mechanical mode. It 
has been investigated for the OM 
\cite{PhysRevLett.96.103901,1367-2630-10-9-095013,omrevasp} as well as for the 
SC system \cite{JuliaPhononLaser,*PhysRevB.88.064305}. Here, we focus on the comparison of the two 
systems.\\
Induced phonon creation, that is exploited for phonon lasing can be 
understood as an anti-Stokes process of at least second order. For a most efficient 
build up of phonon population, the pump laser has to be blue detuned with 
respect to the HF resonance by roughly a multiple of the phonon frequency $n\Omega$. 
Due to a higher order energy conserving process, an excitation of the LF 
mode is achieved, when exciting the HF component [cf. Fig. 
\ref{fig:Schemaheat}]. At this point a major difference between the OM and the 
SC system already becomes apparent: For the OM system the
upper- and the lower level in Fig. \ref{fig:Schemaheat} may represent
two arbitrary photon number states with $n-1$ and $n$ photons. Thus, the
process is not restricted by the cavity decay rate 
$\kappa$. However, for the SC system, the excitation process can only occur, if 
the QD is decayed into its ground state.
Therefore, the possibility of creating 
a phonon within this anti-Stokes process is limited by the decay time of the 
QD excitation.\\
In the semiclassical model phonon lasing manifests itself as a parametrical 
instability: The steady state of low phonon number [cf. Sec. \ref{sec:bistab}] 
becomes unstable and the system reaches a periodic orbit of high phonon number. 
This is shown in Fig. \ref{fig:LasingEvoIns} for the OM system and the SC system 
with LA and LO phonons. For the OM and the SC system with LA phonons
the amplitude of the oscillations is small in comparison to the number of 
phonons since the damping of the phonon mode is rather small [See tab. 
\ref{tab:damppars}]. This way, new phonons are generated via the above described 
process, while phononic excitation is still present. However, in the SC 
system with LO phonons, the phonon lifetimes are typically very small 
\cite{}, while the radiative lifetime of the QD excitations remains 
the same.
The creation of LO-phonons, in particular coherent phonons is therefore not as 
efficient [Fig. \ref{fig:Schemaheat}] and the phononic excitation decays 
for the most part before a new phonon is created.\\
In Figs. \ref{fig:Lasingdensity} we show the mean phonon number in the system as a
function of the detuning and the pump strength for the different systems. In 
case of the SC system, the number of QDs is further illustrated to have a 
crucial impact on the phonon lasing properties, such as thresholds or maximally 
achieved phonon number.
For the OM system, we observe a maximum of the mean phonon number at the 
phonon assisted resonance for lower pump strengths [cf. Figs. \ref{fig:Lasingdensity}(a)].
Pumping more strongly involves higher order processes, leading to a maximum at higher detunings.
For the SC system, we note first of all, that the maximum phonon number is roughly
proportional to the number of QDs [cf. Figs. \ref{fig:Lasingdensity}(b)-(d)], which is consistent 
with the fact, that the QD excitations have to decay before the generation of 
a phonon can take place, i.e. in order to realize a phonon laser cycle. The 
next aspect, that can be observed, is that the maxima for phonon lasing are at 
the polaron-shifted phonon assisted resonances for lower pump strengths: In the 
case of a single QD $N=1$, only the first phonon-assisted resonance involving 
the first order anti-Stokes process (as depicted in Fig. \ref{fig:Schemaheat}) 
contributes to phonon lasing. For a larger number $N$ of QDs [cf. Figs. 
\ref{fig:Lasingdensity}(c),(d)] also the resonance involving two phonons becomes visible. In 
relation to this, we observe a shift of the resonance-condition for phonon-lasing. This can be evaluated 
approximately by considering an effective system accounting for the second order 
process depicted above. This is shown in App. \ref{app:effectivesystem}, which follows Ref. \cite{JuliaPhononLaser,*PhysRevB.88.064305}.
The detuning from the effective QD resonance can be approximated as 
$\Delta_{\mathrm{eff}}=-\Delta-2\frac{E_1^2}{\Delta}-\nHF\frac{g^2}{\Omega}$. The resonance given by the detuning (first term), as defined in \ref{sec:comp}, is altered by self-quenching 
(second term) with the pump strength and by the polaron shift (third term) with 
the electron-phonon coupling strength. For more than one QD [cf. Figs. 
\ref{fig:Lasingdensity}(c),(d)] the polaron shift is proportional to the number of QDs.
However, the QD excitation saturates fast, so that the shift is approximately constant in the lasing
regime. The characteristic shift with the pump strength originates in the case of the SC system
from the self quenching.\\
For the OM system, the effective detuning is $\Delta_{\mathrm{eff}}=-\Delta-\frac{g^2}{\Omega}\nHF$.
This reveals, that there is no self-quenching, but also a dispersive shift,
which is proportional to the number of photons in the cavity. This is analog 
to the proportionality to $\nHF$ in the SC case. Due to the weak coupling strength 
for most OM systems, the dispersive shift becomes only important for very high 
pump strengths.\\
Further, for larger numbers of QDs [cf. Fig. \ref{fig:Lasingdensity}(d)], the one and two 
phonon lines mix so that the maximum for phonon lasing not only involves a 
first, but also a second order phonon-assisted process. Then the above 
approximation becomes invalid.\\
The behavior of the SC system with LO phonons is similar to the one with LA 
phonons, but with lower phonon numbers due to the inefficient creation of 
phonons discussed earlier [cf. Fig. \ref{fig:Lasingdensity}(e),(f)]. Nonetheless, the phonon number 
is still approximately proportional to the number of QDs, so that higher phonon numbers can 
be achieved by increasing the number of QDs. The mean number of phonons excited 
by the second order process is in a similar order of magnitude as the number of 
phonons in the stationary state for direct excitation. As in the case of LA 
phonons, the polaron shift and the self-quenching can be observed as well.

\section{Enhancement of damping}
\label{sec:damping}

\begin{figure}
  \centering
  \includegraphics[width=0.6\linewidth]{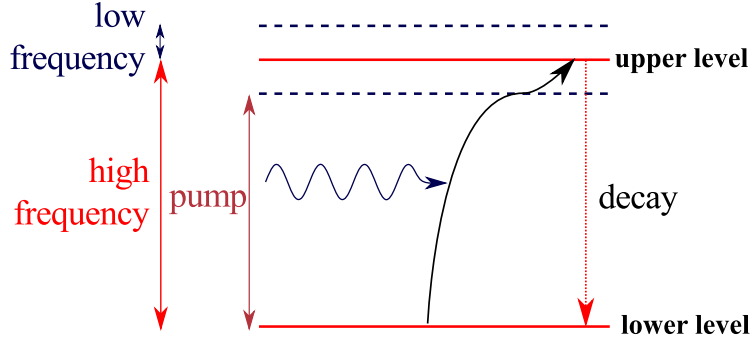}
  \caption{Scheme of the mechanism leading to enhanced damping. See text for 
explanation.}
  \label{fig:Schemacool}
\end{figure}

The last effect, we will consider is the enhancement of the damping of the LF 
component through the interaction with the HF component. This effect is closely related 
to the lasing, which we examined in the previous section. It is also the basic mechanism 
involved in optomechanical backaction cooling.
\cite{PhysRevLett.99.093901,PhysRevLett.99.093902,PhysRevA.77.033804}
For a full description of cooling, thermal fluctuations have to be taken into 
account,
which will not be done in this manuscript. Here, we employ a simplified analysis and 
set the
focus on the enhancement of the damping, which is usually treated in terms of a 
linear
response analysis of the system of Eq. \eqref{eq:eom}. \cite{omrevasp,PhysRevA.77.033804}\\
The effective damping rate of the LF component, which gives the time the system 
needs to return to its stationary state, can be altered by the interaction with the 
HF component.
At first we will investigate the increase of the damping in the OM and the SC 
system
by evaluating the largest Lyapunov exponent for the LF part of the linearized 
equations of motion, as computed for the stability analysis in Sec. \ref{sec:bistab}.
\cite{PaperOptoControl2014}
This enables us to determine the most persistent oscillations in the phononic 
subsystem.\\
\begin{figure}
  \centering
  \includegraphics[width=0.7\linewidth]{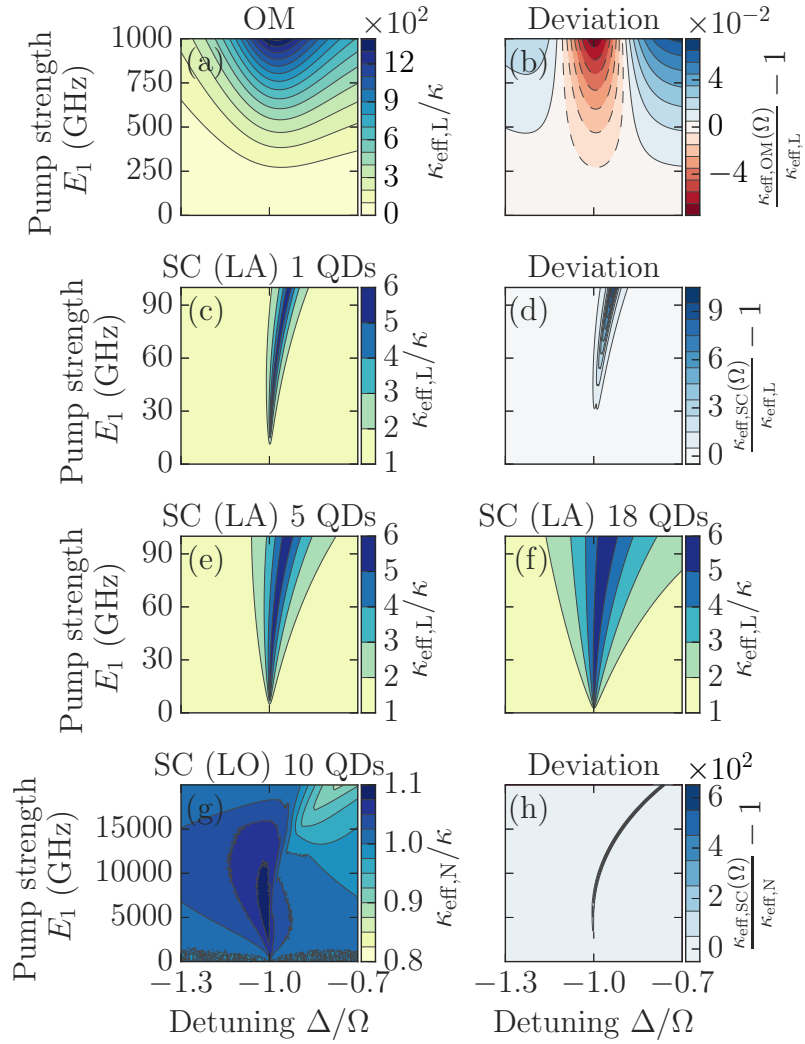}
  \caption{Here, the enhancement of the damping for the different systems as a function of pump strength and detuning is shown. (a) The damping in the OM system is enhanced by a factor of up to 1400 with the chosen parameters. (b) The analytical formula \eqref{eq:kappaeffOM} is in good agreement in this case, since the deviation is only in the order of $10^{-2}$. (c,e,f) For increasing numbers of QDs, the width of the enhancement of the damping increases, but it saturates at a factor of about 6. The analytical formula (d) predicts an enhancement with additional pump and with the number of QDs [cf. \eqref{eq:kappaeffSC}]. For the SC system with LO phonons, the damping cannot be enhanced significantly (g). Even though the analytical formula only predicts a very high enhancement (h).}
  \label{fig:Coolingdensity}
\end{figure}
In Figs. \ref{fig:Coolingdensity}, the largest Lyapunov exponent $\kappa_{\mathrm{eff,L}}$ is shown for the different systems as
a function of the detuning $\Delta$ and the pump strength $E_1$. Note, that for the
cooling in the OM case different parameters than for the bistability in Sec.
\ref{sec:bistab} are used. This is done, because the stationary states become
unstable shortly after the bistability for the parameters in Sec. 
\ref{sec:bistab}
and the enhancement of the damping for the stable states is rather small in 
this 
case.
To avoid this, a set of parameters with weaker coupling is used [cf. Tab. 
\ref{tab:damppars}].
For the SC system it is possible to use the same parameters as in the case of 
the
bistability, since the damping saturates before the relevant stationary state 
becomes unstable.\\
For the OM case [cf. Fig. \ref{fig:Coolingdensity}(a)], we observe an enhancement of the
damping with increasing pumping power up to a factor of 1400. 
In the SC case with LA phonons [cf. Fig. \ref{fig:Coolingdensity}(c,e,f)], the damping is also 
enhanced but only
by a factor of about six. This is to be expected, as the cooling utilizes the 
process complementary to lasing [cf. Fig. \ref{fig:Schemacool}]: When the HF component 
is pumped, due to the detuning in the order of the negative phonon frequency, a phonon is 
absorbed. This process can again only be as fast as the decay of the QD excitation, while 
this restriction is not present in the OM case.
In contrast to the lasing case, where more QDs could increase the maximal 
number of phonons, the damping cannot be increased by adding additional QDs.
This is due to the fact, that the stationary phonon number is rather small in case of
the SC systems ($<1$). This way not enough phonons are present in the system, to make use
of the mechanism depicted in Fig. \ref{fig:Schemacool} with multiple phonons simultaneously.\\
In the SC case with LO phonons, the damping cannot be computed via the largest Lyapunov exponent,
since the phonon damping is not the slowest decay in this case [cf. Tab. \ref{tab:damppars}].
Here, we evaluate the damping numerically by fitting an exponential envelope to the time
evolution of the phonon position. The damping determined this way is called $\kappa_{\mathrm{eff,N}}$.
Fig. \ref{fig:Coolingdensity}(g) shows the damping in the SC system for 10 QDs. The enhancement is so small, since the decay of the QD excitation is the same as in the LA case, while the phonon damping itself is already very high. Thus, the relevant process can only influence the damping marginally.

For small coupling strengths or pump rates, an analytical formula for the 
effective damping
can be derived, as the dynamics in vicinity of the steady state is that of an 
effective
damped oscillator without any interactions: $\dot{B}_{\mathrm{eff}}=-\left(i 
\Omega_\mathrm{eff}+\kappa_{\mathrm{eff}} \right)B$.
Here, we also included an effective frequency $\Omega_\mathrm{eff}$, since the 
coupling in
principle also changes the oscillation frequency. For the parameters used in 
our cases, however, the effective frequency can be approximated by $\Omega$.
Note, that the effective treatment is only valid when the dynamics in vicinity of 
the steady
state can be described by a damped harmonic oscillator. In general more complex 
dynamics are possible so that the analytical formula looses its validity.\\
To derive the formula, an external force acting on the LF component is taken 
into account
in the Eqs. \eqref{eq:eom}. This force could be a stochastic force originating 
from thermal
fluctuations, which is needed for the analysis of cooling. The resulting 
equations are
linearized in vicinity of the equilibrium positions and the result is Fourier 
transformed.
From the resulting set of linear equations a susceptibility as the linear 
response to
the external force can be computed. This is then compared to the susceptibility 
of
a harmonic oscillator to incorporate the effects of the HF component on the LF 
component
in the form of the frequency and the damping rate of an effective damped 
oscillator.
More details on the calculations and all definitions are given in the App. \ref{app:effdampana}.
For the OM system, the effective damping rate is \cite{PhysRevA.77.033804}
\begin{equation}
\kappa_{\mathrm{eff,OM}}\left(\omegabar\right)=\kappa - \frac{4 g^2 C_s^*C_s 
\Omega \gamma \Deltasch}{\left[\gamma^2 + \left( \omegabar + \Deltasch 
\right)^2 
\right] \left[\gamma^2 + \left( \omegabar - \Deltasch \right)^2 \right]},
\label{eq:kappaeffOM}
\end{equation}
which is valid for $C^*C \gg 1$. The formula for the SC case reads
\begin{widetext}
\begin{multline}
\kappa_{\mathrm{eff,SC}}\left(\omegabar\right)= \kappa - \frac{2 N g^2 E_1 
\Omega}{\omegabar^2+4\gamma^2}\\
\left\lbrace \left[ \Re{P_s} \Deltasch- \Im{P_s} \left( \gamma-\gammapd \right) 
\right]\left( \gamma_r^2-\Deltasch^2-\omega_R^2 \right)\right.\\+\left.2\left[ 2 \Re{P_s} 
\Deltasch \gamma + \Im{P_s} \left( 2 \gamma \gammasch + \omegabar^2 \right) 
\right]\gamma_r\sqrt{1+\frac{\mathcal{D}}{2}} \sqrt{1-\frac{4 
E_1^2}{\omegabar^2+4\gamma^2}} \right\rbrace \\/\left\lbrace  \left[ \gamma_r^2+\left( \omega_R+\Deltasch 
\right)^2 \right] \left[ \gamma_r^2+\left( \omega_R-\Deltasch \right)^2 \right] 
+2 \mathcal{D} \gamma_r^2\omega_R^2 \right\rbrace 
\label{eq:kappaeffSC}
\end{multline}
\end{widetext}
and is valid for weak excitation.\\
In both cases $\Deltasch=\Delta+\frac{N g^2 \Omega U_s}{\Omega^2+\kappa^2}$.
This includes the dispersive shift in case of the stationary state. This reveals
why the dispersive shift is only observed in the OM system: The dispersive shift
is proportional to the stationary number of excitations in the system, which is 
small
for the SC system, while it is significant in the OM case. Even though the 
formula for the SC case is more complicated some similarities can be observed. In case 
of weak excitation the formula can be simplified by approximating $\omega_R 
\approx 
\omega$, $\gamma_r\approx \gammasch$,
and $\mathcal{D}\approx 0$. The major restriction of the analytical formula is,
that it relies on harmonic dynamics. For both systems this approximation fails
at some point, but since the coupling in the OM system is weaker, the region of
validity is very large and the deviation is only small. This is shown in Fig. \ref{fig:Coolingdensity}(b) for the OM and in Fig. \ref{fig:Coolingdensity}(d) for the SC 
system. The deviation of the analytical formula from the full computation is 
given: For the OM case, only a small differences in the order of a few percent are observed [cf. Fig. \ref{fig:Coolingdensity}(b)],
while for the SC systems, the formula overestimates the effect of cooling drastically, especially for large pump powers [cf. Figs. \ref{fig:Coolingdensity}(f,h)].

%\clearpage

\section{Conclusion}

In conclusion, we studied similarities in the behavior of an optomechanical and a semiconductor
system with a single phonon mode, constituted by either a single acoustic phonon mode or
an effective optical phonon mode. We showed that the mechanical mode of the mirror has the same function
as the phonon mode. Also, the cavity resonance and the optical resonance of the quantum dot are analogous.
The underlying model is a Hamiltonian formulation, from which equations of motion are derived.\\
Even though the systems exhibit similar features, the underlying physics are 
different. In particular, as already discussed in the model section \ref{sec:comp}, the 
high frequency component is bosonic in the optomechanical case, while it is fermionic in the semiconductor case.\\
Both systems exhibit the nonlinear effect of bistability. Here, the optomechanical system can be understood as
the limit of the semiconductor system with many quantum dots. This is the case, since the additional equation, needed
for the description of the fermionic quantum dots, is not necessary for weak excitation of the quantum dots.\\
Furthermore, the detuning of the system resonance from the pump laser can be used to
control the behavior of the phonon mode: In the blue-detuned regime, the number of
phonons is strongly increased, leading to phonon lasing. In the red-detuned regime,
the damping of the phonon mode is enhanced.\\
In case of phonon lasing, the number of coherent phonons is very large for the optomechanical system
since the relevant processes are not limited strictly by the lifetime of the high frequency component. For the 
semiconductor system, this limitation is given, since the quantum dot can only be excited again after its
excitation has decayed. 
The phonon number can be increased by additional quantum dots coupled to the phonon mode for the semiconductor system.
In case of optical phonons, however, their short lifetimes prevent a significant occupation
of the phonon mode.\\
The damping may be enhanced for the optomechanical system and the semiconductor system with acoustic phonons.
For optical phonons an enhancement of the damping cannot be observed. This is attributed
to the choice of parameters, especially to the short phonon lifetime.
For the optomechanical system, the damping may be increased strongly before other effects dominate and prevent
an enhancement. For the semiconductor system, the damping may only be enhanced weakly. Even additional quantum dots
do not allow a further enhancement, which would be suggested from the observations for lasing.

\section*{Acknowledgments}
We are grateful towards the Deutsche Forschungsgemeinschaft for support
through SFB 910 and SFB 787 as well as the Sandia LDRD
program, funded by the U.S. Department of Energy under
Contract No. DE-AC04-94AL85000.

%Since the  optomechanical coupling is generally smaller than the coupling of the
%QD to the acoustic phonon mode, the bistability occurs at a larger pump power with an
%overall higher number of excitations
%in the system. This is also the case for the enhancement of the damping, which 
%can only be observed at a high intracavity power. Thus, devices exploiting the effects in SC
%systems would consume less power. Nonetheless, ground state cooling of mechanical modes
%is difficult to achieve with in SC systems, because, even though we chose similar
%quality factors for both systems in the study at hand, the presently achievable
%quality factors ($Q=\omega/\kappa$) for OM cavities are higher, than for phonon
%cavities \cite{}. Additionally, even though the optomechanical coupling is 
%generally smaller, the enhancement of the damping works still well in OM 
%systems,
%since the effective coupling is increased with the photon number and with this
%also the damping. This can be seen from Eqs. \eqref{eq:kappaeffOM} and\eqref{eq:kappaeffSC}.

\appendix

\section{Limiting behavior of the bistability}
\label{app:NQDlimit}
In case of the bistability [cf. Sec. \ref{sec:bistab}] the necessity for an equilibrium of forces $F_h+F_c=0$ leads to a cubic equation with up to three real roots in the bistable regime. When using the relation $q_s=-\frac{\sqrt{2} g \Omega U_{\mathrm{tot}}}{\Omega^2+\kappa^2}$, which follows from \eqref{eq:eomb} in the stationary limit, the cubic equation can be expressed in terms of the excitation of the HF component. Note, that $U_{\mathrm{tot}}=P^*P$ in the OM case and $U_{\mathrm{tot}}=N U$ in the SC case. For the OM system, the equation, which has to be solved reads
\begin{equation}
\left[ \left( \Delta+\frac{2 g^2 \Omega U_{\mathrm{tot}}}{\Omega^2+\kappa^2} \right)^2+\left( \gamma+\gammapd \right)^2 \right] U_{\mathrm{tot}}=E_1^2.
\label{eq:OMcubic}
\end{equation}
The corresponding equation for the SC system reads
\begin{equation}
\left[ \left( \Delta+\frac{2 g^2 \Omega U_{\mathrm{tot}}}{\Omega^2+\kappa^2} \right)^2+\left( \gamma+\gammapd \right)^2 \right] U_{\mathrm{tot}}=N E_1^2\left( 1 - \frac{ \gammapd}{\gamma} \right)\left(1-2 \frac{U_{\mathrm{tot}}}{N} \right).
\label{eq:SCcubic}
\end{equation}
If we neglect pure dephasing and assume $\frac{U_{\mathrm{tot}}}{N}\ll 1$, i.e. a single QD is only excited weakly, the rule discussed in Sec. \ref{sec:bistab} for connecting the systems can be observed by comparing Eqs. \eqref{eq:OMcubic} and \eqref{eq:SCcubic}. 

\section{Effective system}
\label{app:effectivesystem}

To gain further insight into the processes, which are relevant for lasing, an effective Hamiltonian is derived \cite{JuliaPhononLaser,*PhysRevB.88.064305}. This takes only the second order process discussed in Sec. \ref{sec:lasing} into account [cf. Fig. \ref{fig:Schemaheat}].\\
This is done in two steps: At first, a unitary transformation is applied in order to transform the Hamiltonian into a rotating frame. Afterwards, another unitary transformation is used to eliminate interactions in first order. The rotating frame is introduced by
\begin{align}
H_{\mathrm{RF}}&=U H U^{\dagger} - i\hbar U \partial_t U^{\dagger},\\
U&=e^{\frac{i}{\hbar}\xi t},
\nonumber\\
\xi&= \hbar \omegaL \pd\pv.
\nonumber
\end{align}
This leads to the Hamiltonian
\begin{equation}
H_{\mathrm{RF}}=\underbrace{\hbar\Omega\bd\bv-\hbar\Delta\pd\pv}_{=H_0}+\underbrace{\hbar g \pd\pv\left(\bv+\bd \right)+i\hbar\mathbf{E}\left(\pd-\pv\right)}_{=H_{\mathrm{I}}},
\end{equation}
where, as above, $\Delta=\omegaL-\omega$. This way, the explicit time dependence of the Hamiltonian is eliminated. In the next step, the unitary transformation
\begin{multline}
H_{\mathrm{eff}}=e^{i S} H e^{-i S} \approx H_{0} + 
H_{\mathrm{I}} 
+ [i S,H_{0} + H_{\mathrm{I}}]  + \frac{1}{2} [i S,[i 
S, H_{0}]]
\label{eq:Heffexp}
\end{multline}
is applied such that only interaction terms of second order are present.
By choosing $S=\boldsymbol{\alpha} \pv+\beta \pd\pv\bv+H.c.$, the first order terms can be eliminated, and the effective Hamiltonian can be computed from
\begin{equation}
H_{\mathrm{eff}}=H_{0}+\frac{1}{2}[i S, H_{\mathrm{I}}].
\label{eq:Hefftrans}
\end{equation}
This is achieved with $\boldsymbol{\alpha}=-\frac{\mathbf{E}}{\Delta}$ and $\beta=i\frac{g}{\Omega}$.
The effective Hamiltonian reads then
\begin{multline}
H_{\mathrm{eff}}=
\hbar \Omega \bd\bv-\hbar\Delta\pd\pv+\frac{\hbar}{\Delta} W - \hbar \frac{g^2}{\Omega}\pd\pv\pd\pv\\
+i\hbar\frac{g \mathbf{E}}{2}\left( \frac{1}{\Delta}+\frac{1}{\Omega} \right) \left(\pd \bd-\pv \bv\right)\\
+i\hbar\frac{g \mathbf{E}}{2}\left(\frac{1}{\Delta}-\frac{1}{\Omega}\right) \left(\pd \bv -\pv \bd\right).
\label{eq:Heff}
\end{multline}
Here, we shifted the energy scale appropriately. The first two terms are the energy of
the LF component and of the HF component in the rotating frame. The third term can
be neglected in the OM case, where it constitutes a shift of the total energy scale
of the Hamiltonian ($W_{\mathrm{OM}}= \mathbf{E}^2$). In the SC case, this term
introduces the self-quenching and can be written as
$W_{\mathrm{SC}}=2\sum_{i=1}^{\NQD} E_i^2 \hat{p}_i^{\dagger}\hat{p}_i$.
The fourth term describes the dispersive shift. It depends on the number of excitations in the system.
The fifth term models two processes, which are energy conserving in the case of
lasing [cf. Fig. \ref{fig:Schemaheat}]: The number of excitations in the HF component
is increased (decreased) by the external field and a phonon is emitted (absorbed).
The last term describes processes, which are energy conserving in case of
cooling [cf. Fig. \ref{fig:Schemacool}]: The number of excitations in the HF component
is increased (decreased) by the external field and a phonon is absorbed (emitted).
Depending on the scenario one of the two terms can be neglected.

\section{Effective damping analytical}
\label{app:effdampana}

The formulas for the effective damping are derived in analogy to Ref. \cite{PhysRevA.77.033804}, while we use the semiclassical equations as starting point. For the linearized system, this leads to the same susceptibility \cite{omrevasp}. At first the system of equations \eqref{eq:eom} is linearized in vicinity of the steady state. In addition an external force $F$ on the LF component is taken into account: $\dot B =-(i \Omega +\kappa) B - i N g U + i F/\sqrt{2}$.
Then, the real variables $q=\frac{B+B^*}{\sqrt{2}}$, $p=\frac{B-B^*}{\sqrt{2 i}}$, $X=\frac{P+P^*}{\sqrt{2}}$, and $Y=\frac{P-P^*}{\sqrt{2 i}}$ are introduced and the the deviation from the steady state is defined as $\delta q(t) = q(t)-q_s$. Here, the index $s$ indicates the stationary value. When terms, which are nonlinear in the deviations, are neglected, a set of linear differential equations emerges. This can be Fourier transformed, leading to the sets of equations
\begin{subequations}
\label{eq:OMeomF}
\begin{align}
-i \omegabar \mathcal{F}_q &= \Omega \mathcal{F}_p - \kappa \mathcal{F}_q
\label{eq:OMeomqF} \\
-i\omegabar \mathcal{F}_p &= -\Omega \mathcal{F}_q-\kappa \mathcal{F}_p - g_X \mathcal{F}_X - g_Y \mathcal{F}_Y+ \mathcal{F}_F
\label{eq:OMeompF}\\
-i\omegabar \mathcal{F}_X &= \Deltasch \mathcal{F}_Y- \gamma \mathcal{F}_X + g_Y \mathcal{F}_q
\label{eq:OMeomXF}\\
-i\omegabar \mathcal{F}_Y &= -\Deltasch \mathcal{F}_X- \gamma \mathcal{F}_Y - g_X \mathcal{F}_q
\label{eq:OMeomYF}
\end{align}
\end{subequations}
for the OM system and
\begin{subequations}
\label{eq:SCeomF}
\begin{align}
-i \omegabar \mathcal{F}_q &= \Omega \mathcal{F}_p - \kappa \mathcal{F}_q
\label{eq:SCeomqF} \\
-i\omegabar \mathcal{F}_p &= -\Omega \mathcal{F}_q-\kappa \mathcal{F}_p - \sqrt{2} N g \mathcal{F}_U + \mathcal{F}_F
\label{eq:SCeompF}\\
-i\omegabar \mathcal{F}_X &= \Deltasch \mathcal{F}_Y- \tilde\gamma \mathcal{F}_X + g_Y \mathcal{F}_q - 2\sqrt{2} E_1 \mathcal{F}_U
\label{eq:SCeomXF}\\
-i\omegabar \mathcal{F}_Y &= -\Deltasch \mathcal{F}_X- \tilde\gamma \mathcal{F}_Y - g_X \mathcal{F}_q
\label{eq:SCeomYF}\\
-i\omegabar \mathcal{F}_U &= \sqrt{2} E_1 \mathcal{F}_X - 2 \mathcal{F}_U
\label{eq:SCeomUF}
\end{align}
\end{subequations}
for the SC system. The $\mathcal{F}$ indicates the Fourier transform of the respective variable. The Fourier transforms are always functions of $\omegabar$. We introduced the definitions $\tilde\gamma=\left(\gamma+\gammapd\right)$, $\Deltasch=\Delta+\sqrt{2} g q_s$, $g_X=\sqrt{2} g X_s$, and $g_Y=\sqrt{2} g Y_s$.\\
% Note the two main differences between the systems. In case of the SC system, there is possibly the additional pure dephasing. Furthermore, $\mathcal{F}_U$ influences $\mathcal{F}_p$, and not both \\
This system of equations can be solved so that, the susceptibility as the linear response to the external force can be computed from $\mathcal{F}_q=\chi\left(\omegabar \right)\mathcal{F}_F$. In the case of the OM system, the inverse of the susceptibility reads
\begin{equation}
\chi_{\mathrm{OM}}^{-1}\left(\omegabar\right)=-\frac{\omegabar^2+2 i \omegabar\kappa -\kappa^2+\Omega^2}{\Omega}-\frac{2 g^2 P_s^*P_s \Deltasch}{\left( \gamma-i\omegabar \right)^2+\Deltasch^2}.
\label{eq:chiOM}
\end{equation}
For the SC system, it is
\begin{multline}
\chi_{\mathrm{SC}}^{-1}\left(\omegabar\right)=-\frac{\omegabar^2+2 i \omegabar\kappa -\kappa^2+\Omega^2}{\Omega}  + \frac{2 g E_1 N}{i \omegabar-2\gamma} \frac{g_X \Deltasch+ g_Y \left(i \omegabar-\tilde\gamma \right)}{\left( -i\omega\left(1-R\right) +\tilde\gamma + 2 R \gamma\right)\left( \tilde\gamma-i\omegabar \right)+\Deltasch^2}.
\label{eq:chiSC}
\end{multline}
Here, we introduced the definition $R =\frac{4 E_1^2}{\omegabar^2+4\gamma^2}$. Differences to Ref. \cite{PhysRevA.77.033804} for the OM system arise from differences in the definition of the parameters [cf. Tabs. \ref{tab:bistpars}, \ref{tab:damppars}] and the introduction of damping. In the limit $\Omega\gg\kappa$, which is valid for the systems we are interested in, both descriptions are equivalent.
From the susceptibilities, the effective dampings \eqref{eq:kappaeffOM} and \eqref{eq:kappaeffSC} can be computed, where we use
\begin{align*}
\mathcal{D}&=\frac{1}{2}\frac{R^2\left(1-\frac{2\gamma}{\tilde\gamma} \right)^2}{1-\left( 1+\frac{2\gamma}{\tilde\gamma}\right) R +\frac{2\gamma}{\tilde\gamma} R^2 }\\
\gamma_r&= \tilde\gamma\sqrt{1-\frac{2\gamma}{\tilde\gamma} R}\\
\omega_R&= \omegabar \sqrt{1+R}.
\end{align*}

\section{Parameters}
\label{app:pars}

\begin{table}[h!]
	\caption{\label{tab:bistpars} Parameter values for the bistabilities in 
Sec. \ref{sec:bistab}.}
	\begin{tabular}{ c | c | c | c }
	%\hline
	 & Optomechanical \cite{qmombistab} & Semiconductor (LA) \cite{JuliaPhononLaser,*PhysRevB.88.064305} & Semiconductor (LO) \\
	\hline
	\hline
	Detuning of pump and & $\Delta = -2.6\times\Omega$ & $ \Delta = -\Omega/8$ & $\Delta=-\Omega$
\\
	%\hline
	LF component & $\Omega= 2 \pi\times 10 \text{ MHz}$ & 
$\Omega = 556.6 \text{ GHz}$ & $\Omega = 55.3 \text{ THz}$ \\
	%\hline
	Losses & opt. cavity $\gamma = 2 \pi\times 14 \text{ MHz}$ & TLS 
$\gamma = 5 \text{ GHz}$ & $\gamma = 5 \text{ GHz}$  \\
	& mech. resonator $\kappa =2 \pi\times 50 \text{ Hz}$ & acoust. 
cav. 
$\kappa = 0.5 \text{ GHz}$& $\kappa = 100 \text{ GHz}$\\
 & & & $\gammapd = 100 \text{ GHz}$  \cite{PhysRevLett.96.140405}\\
	%\hline
	coupling strength &  $|g| = 952.7 \text{ Hz} $ & $|g|= 197.5 \text{ 
GHz}$&  $|g|= 5.1 \text{ THz}$\\
	%\hline
	Number of QDs & -- & 1 & 10 \\
	%\hline
	\end{tabular}
\end{table}

\begin{sidewaystable}[h!]
	\caption{\label{tab:damppars} Parameter values for enhancing or 
decreasing the damping used throughout Secs. \ref{sec:lasing} and \ref{sec:damping}.}
	\begin{tabular}{ c | c | c | c }
	%\hline
	 & Optomechanical \cite{PhysRevA.77.033804} & Semiconductor \cite{JuliaPhononLaser,*PhysRevB.88.064305} & Semiconductor (LO)\\
	\hline
	\hline
	pump rate & $E_1 = 60 \text{ GHz}$ & $E_1 = 80 \text{ GHz}$ & $E_1 = 9.01 \text{ THz}$ \\
	& [cf. Fig. \ref{fig:LasingEvoIns}(a)] & [cf. Fig. \ref{fig:LasingEvoIns}(b)] & [cf. Fig. \ref{fig:LasingEvoIns}(c)] \\
	%\hline
	Detuning of pump  & $\Delta \approx \pm\Omega$ & $ \Delta \approx 
\pm\Omega$ & $\Delta \approx 
\pm\Omega$\\
	($+$: Lasing, $-$: enhancement of damping) &  &  & \\
	%\hline
	LF component & $\Omega= 2 \pi\times 10 \text{ MHz}$ & 
$\Omega = 556.6 \text{ GHz}$ & $\Omega = 55.3 \text{ THz}$ \\
	%\hline
	Losses & opt. cavity $\gamma = 2\pi\times 2 \text{ MHz}$ & TLS 
$\gamma = 5 \text{ GHz}$ & $\gamma = 5 \text{ GHz}$ \\
 & & & $\gammapd = 100 \text{ GHz}$ \\
	& mech. resonator $\kappa_{\mathrm{m}}= 2 \pi\times 50 \text{ Hz}$ & acoust. cav. 
$\kappa = 0.5 \text{ GHz}$& $\kappa = 50 \text{ GHz}$ \\
	%\hline
	coupling strength &  $|g| = 205 \text{ Hz} $ & $|g|= 197.5 \text{ 
	GHz}$&  $|g|= 5.1 \text{ THz}$\\
	%\hline
	Number of QDs & -- & 1 [cf. Fig. \ref{fig:LasingEvoIns}(b)], 5, 18 &  10 [cf. Fig. \ref{fig:LasingEvoIns}(c)], 50 \\
	%\hline
	\end{tabular}
\end{sidewaystable}
\bibliographystyle{apsrev4-1}
\bibliography{./bibms}

\end{document}